\documentclass{PoS}
\newcommand\T{\rule{0pt}{1.6ex}}  

\title{MAGIC observations on Pulsar Wind Nebulae around high spin-down power \textit{Fermi}-LAT pulsars}

\ShortTitle{MAGIC observations on PWNe around high spin-down power pulsars}

\author{\speaker{A.~Fern\'andez-Barral}\\
        Institut de Fisica d'Altes Energies (IFAE), The Barcelona Institute of Science and Technology (BIST)\\ 
        Campus UAB, 08193 Bellaterra (Barcelona), Spain\\
        E-mail: \email{afernandez@ifae.es}}

\author{O.~Blanch\\
		Institut de Fisica d'Altes Energies (IFAE), The Barcelona Institute of Science and Technology (BIST)\\ 
        Campus UAB, 08193 Bellaterra (Barcelona), Spain\\}
        
\author{A.~Chatterjee\\
		Saha Institute of Nuclear Physics, HBNI\\
		 1/AF Bidhannagar, Salt Lake, Sector-1, Kolkata 700064, India\\}

\author{E.~de O\~na Wilhelmi\\
		Institute for Space Sciences (CSIC/IEEC)\\
		E-08193 Barcelona, Spain\\}
		
\author{D.~Fidalgo\\
		Universidad Complutense de Madrid\\
		E-28040 Madrid, Spain\\}
		
\abstract{Pulsar Wind Nebulae (PWNe) represent the most numerous population of TeV sources in our galaxy. These sources, some of which emit very-high-energy (VHE) gamma-rays, are believed to be related to the young and energetic pulsars that power highly magnetized nebulae (a few $\mu$G to a few hundred $\mu$G). In this scenario, particles are accelerated to VHE along their expansion into the pulsar surroundings, or at the shocks produced in collisions of the winds with the surrounding medium. Those energetic pulsars can be traced using observations with the Fermi-LAT detector. 
The MAGIC Collaboration has carried out deep observations of PWNe around high spin-down power Fermi pulsars. We study the PWN features in the context of the already known TeV PWNe. We present here the analysis accomplished with three selected PWNe: PSR J0631+1036, PSR J1954+2838 and PSR J1958+2845. }

\FullConference{35th International Cosmic Ray Conference  -ICRC217-\\
		10-20 July, 2017\\
		Bexco, Busan, Korea}

\begin{document}

\section{Introduction}
Pulsars, highly magnetized rotating neutron stars, are constantly releasing their rotational energy in the form of relativistic Poynting and particles flux, the so-called \textit{pulsar wind}. This wind interacts with the interstellar medium (ISM), giving rise to a termination shock in which particles are accelerated. When flowing out, the relativistic particles can, in turn, interact with the surrounding medium generating a magnetized bubble known as Pulsar Wind Nebula (PWN). For the first thousand years, emission from this nebula is mainly synchrotron dominated and detected from radio to X-rays. In the gamma-ray regime, emission is produced through inverse Compton (IC) up-scattering of low-energy photons, composed by CMB, far infrared (FIR) or near infrared (NIR) and optical photons \cite{GaenslerSlane2006}. Based on observational criteria, TeV PWNe are expected when hosting high-spin down power pulsars ($\gtrsim 10^{34}$ erg s$^{-1}$) (see e.g. \cite{StefanPWNstudy}).

PWNe correspond to the most numerous galactic very-high-energy (VHE; $E>100$ GeV) population in the Milky Way. The MAGIC telescopes deeply studied the most luminous one, the Crab Nebula \cite{MAGIC_Crab}, and discovered the least luminous PWN, 3C 58 \cite{Ruben_3C58}. The southern hemisphere Imaging Atmospheric Cherenkov Telescopes (IACTs) H.E.S.S. extensively studied this type of source as well, and were able to firmly identify 14 PWNe during their Galactic Plane Survey (HGPS; \cite{StefanPWNstudy}) towards the inner part of our Galaxy. In this work, we aim to prove particle acceleration at the outer side of the Galaxy, for which we selected promising PWN candidates based on the high spin-down power (between $\sim 10^{35}$--$10^{37}$ erg s$^{-1}$) and characteristic age (few tens of kyr) of the hosted \textit{Fermi}-LAT pulsars: PSR J0631+1036, PSR J1954+2838 and PSR J1958+2845. Basic information from these pulsars can be found in Table \ref{tab:PSR_features}, taken from the ATNF pulsar catalogue\footnote{http://www.atnf.csiro.au/research/pulsar/psrcat/} \cite{ATNF_catalog}. If no distance information is available, this parameter is taken from the literature. The \textit{pseudo-distance} is also provided, which is estimated making use of the spin-down energy loss rate and the gamma-ray luminosity \cite{PseudoDistance}.

\begin{table*}[ht!]
\begin{center}
\caption{Characteristics of the selected PWN candidates for the study. \textit{From left to right:} Spin-down power, characteristic age, distance and pseudo-distance. Information taken from the ATNF catalog if not specified otherwise. $\dot{E}/Distance$ is computed using the so-called Distance.
\footnotesize{$^{a}$ \cite{TianLeahy2006}, $^{b}$ \cite{PseudoDistance}}.} 
\label{tab:PSR_features}      
	\begin{tabular}{ c  c  c  c  c c}
		\hline
		\hline

		Name & $\dot{E}$ & Characteristic Age & Distance & Pseudo-distance & $\dot{E}/\textrm{d}^{2}$ \T \\
		& [erg s$^{-1}$]  & [kyr]  & [kpc] & [kpc] & [erg kpc$^{-2}$ s$^{-1}$] \\
	     \multicolumn{1}{c}{PSR J0631+1036} & $1.7 \times 10^{35}$ & 43.6 & 2.10 & --  & $3.9 \times 10^{34}$ \T  \\
	     \multicolumn{1}{c}{PSR J1954+2838} & $1.0 \times 10^{36}$ & 69.4 & 9.2$^{a}$ & 1.6$^{b}$ & $1.18 \times 10^{34}$ \T \\
	     \multicolumn{1}{c}{PSR J1958+2845} & $3.4 \times 10^{35}$ & 21.7  & 9.2$^{a}$ & -- & $4.0 \times 10^{33}$ \T \\
		\hline	
	\end{tabular}
	\end{center}
\end{table*}

\section{Observations and data analysis}
MAGIC is a stereoscopic system consisting of two 17 m diameter IACTs situated in El Roque de los Muchachos observatory in the Canary island of La Palma, Spain ($28.8^{\circ}$N, $17.8^{\circ}$ W, 2225 m a.s.l.). After a major upgrade that involved digital trigger, readout system and MAGIC\,I camera, the integral sensitivity in stereoscopic mode achieved at low zenith angles is  $0.66\pm0.03$\% of the Crab Nebula flux in 50 hr above 220 GeV \cite{Performancepaper}.\\

Data analysis was performed making use of the MAGIC standard software (MARS, \cite{Zanin2013}). The significance is calculated using Equation 17 from \cite{LiMa1983}, while upper limits (ULs) are obtained following the Rolke method \cite{Rolke2005}  at a 95\% confidence level (CL), assuming a Gaussian background and 30\% systematics on the effective area. All observations presented in this proceeding were carried out under different moonlight conditions, divided according to the $DC$ level during observations. Thus, data was classified as \textit{dark} (if $DC<2.0~\mu$A), \textit{moderate moon} ($2.0~\mu$A$<DC<4.0~\mu$A) and \textit{decent moon} ($4.0~\mu$A$<DC<8.0~\mu$A). The cleaning levels applied in each case correspond to those of $1\times NSB_{\textrm{dark}}$, 2--3$\times NSB_{\textrm{dark}}$ and 5--8$\times NSB_{\textrm{dark}}$ (with nominal HV), respectively, presented in \cite{MoonPerformance}. 

\subsection{PSR J0631+1036}
PSR J0631+1036 belongs to the population of radio-loud gamma-ray pulsars, first detected in a radio search for counterparts of X-ray sources found in Einstein IPC images. VERITAS observed this source in their first year of operation in 2007 for 13 hours, finding no significant excess. Following the Fermi's Bright Source List, eight years of data from the ground based water Cherenkov observatory Milagro were re-analyzed giving rise to a 3.7$\sigma$ hotspot at the position of this source \cite{MilagroonBrigthFermisources}. Nevertheless,  in the recently published second HAWC catalog the hint of possible TeV emission from the direction of PSR J0631+1036 could not be confirmed \cite{2HAWCcatalog}.

MAGIC conducted deep observations of this source for a total of 37 hours during the winter seasons of 2014/15 and 2015/16, within a zenith range of 15 to 50$^{\circ}$ and under different moon conditions.

\subsection{PSR J1954+2838 and PSR J1958+2845}
These two pulsars, reported in the First \textit{Fermi}-LAT Source Catalog (1FGL) \cite{1FGL}, are located in a very dense and crowded region, in which the association of different structures within the field of view is still under debate. PSR J1954+2838 is positionally coincident with SNR G65.1+06, which is a very faint supernova remnant (SNR) at a distance of 9.2 kpc with an estimated age between 40--140 kyr \cite{TianLeahy2006}. This SNR seems to be associated with another pulsar in the FoV, PSR J1957+2831. At the south of the remnant, an IR source is detected, IRAS 19520+2759, which was found to be related to CO, H$_{2}$O and OH emission lines at a distance similar to SNR G65.1+06, suggesting interaction with molecular clouds. The re-analysis of 8 years of Milagro data revealed hot spots at the level of 4.3$\sigma$ and $4.0\sigma$ in the direction of PSR J1954+2838 and PSR J1958+2845, respectively \cite{MilagroonBrigthFermisources}. This emission may originate from the corresponding PWN or interaction of the SNR and molecular cloud, in the case of  PSR J1954+2838. In 2010, MAGIC observed these two pulsars in the stand-alone mode with MAGIC I for $\sim 25$ hours, resulting in a non-detection \cite{MAGICJ1954J1958_2010}. Nevertheless, the major upgrade between 2011--2012 that both telescopes underwent allowed to improve MAGIC sensitivity with respect to former observations. 

In the new campaign, MAGIC observed PSR J1954+2838  for a total of $\sim 16$ hours between April and November 2015, in a zenith range between 5$^{\circ}$ to 50$^{\circ}$. In the case of PSR J1958+2845, only moon data were available, amounting  $\sim 4$ hours of good quality data, in a zenith range of 10$^{\circ}$ to 40$^{\circ}$.

\section{Results}
\subsection{PSR J0631+1036}
We did not find any significant excess in the direction of PSR J061+1036 after $\sim 37$ hours. Following Table A.2. from \cite{StefanPWNstudy}, this source is expected to be extended for MAGIC, taken into account its characteristic age and relative proximity to our solar system. Nevertheless, since its extension was not confirmed observationally so far, we computed ULs for both point-like and disk profile with 0.3$^{\circ}$ radius (maximum possible extension given our observational settings). The corresponding integral ULs above 300 GeV are $6.0 \times 10^{-13}$ cm$^{-2}$ s$^{-1}$ and $2.8 \times 10^{-12}$ cm$^{-2}$ s$^{-1}$, respectively. Both ULs are not in agreement with the Milagro measurement, assuming a power-law spectrum with photon index $\Gamma=2.2$, which corroborates the non-detection reported by HAWC \cite{2HAWCcatalog}. For the following discussion we will adopt the more conservative limit, but note that a larger extension than 0.3$^{\circ}$ could affect our background estimation and render the ULs too optimistic.

\subsection{PSR J1954+2838 and PSR J1958+2845}
\begin{center}
\begin{figure}
\centering
		\includegraphics[width=.5\linewidth]{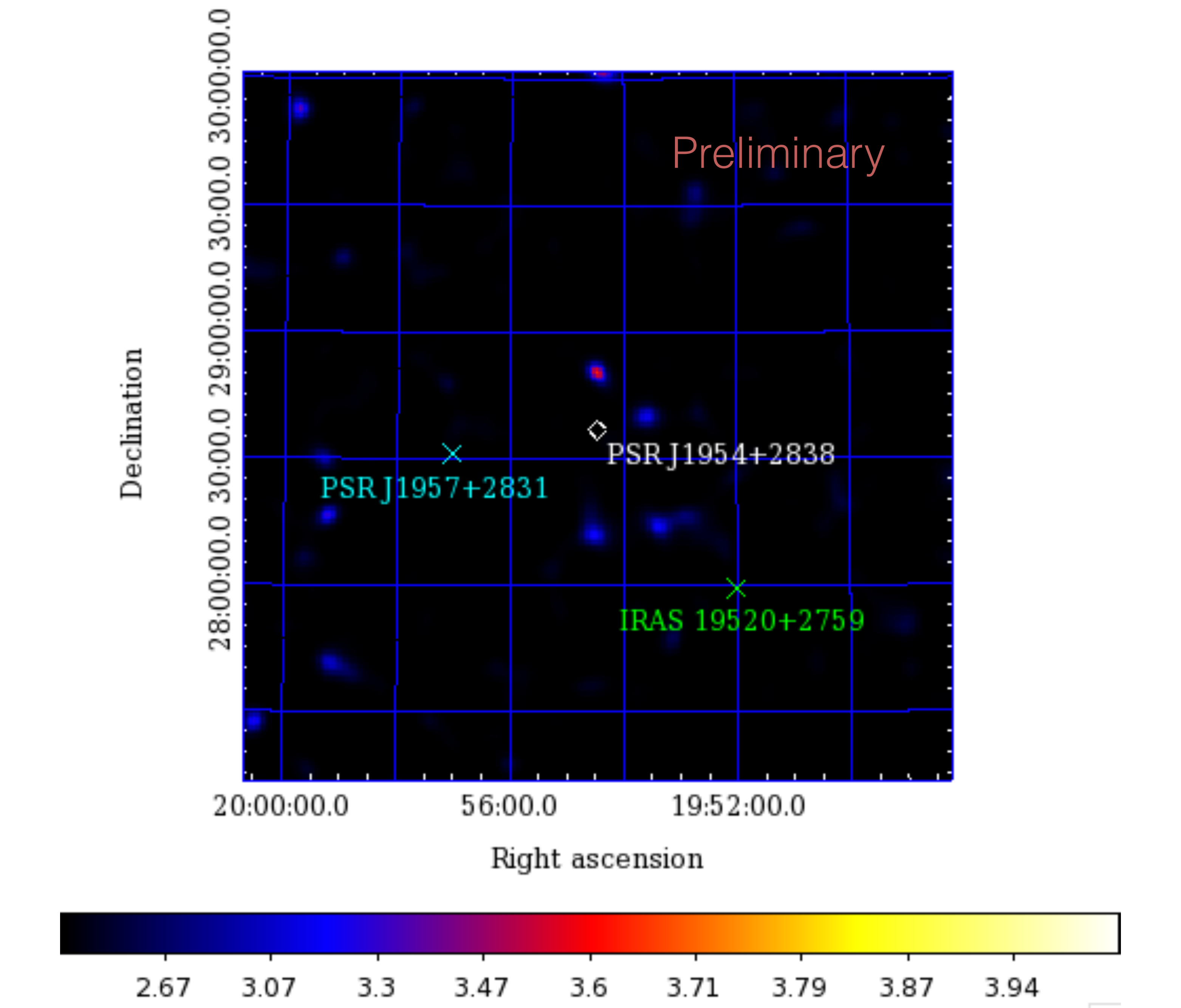}
		\caption{MAGIC significance skymap for the observations of PSR J1954+2838 (white diamond). The pulsar PSR J1957+2831 associated to the SNR G65.1+06 is marked in blue, while the IR source, IRAS 19520+2759, located at the south of the remnant, is shown in green.}
		\label{fig:skymapJ1954}
\end{figure}
\end{center}

No gamma-ray excess was found in the direction of either PSR J1954+2838 or PSR J1958+2845. The measured signal is compatible with background at energies greater than 300 GeV and 1 TeV (the latter motivated by Milagro hotspots). A hotspot situated in the FoV of PSR J1954+2838 appeared at an offset of $\sim 0.23^{\circ}$ from the nominal source at the level of $\sim 3.5\sigma$, although its position is not coincident with any known system (see Figure \ref{fig:skymapJ1954}). 

\begin{center}
\begin{figure*}
\centering
		\includegraphics[width=1.\linewidth]{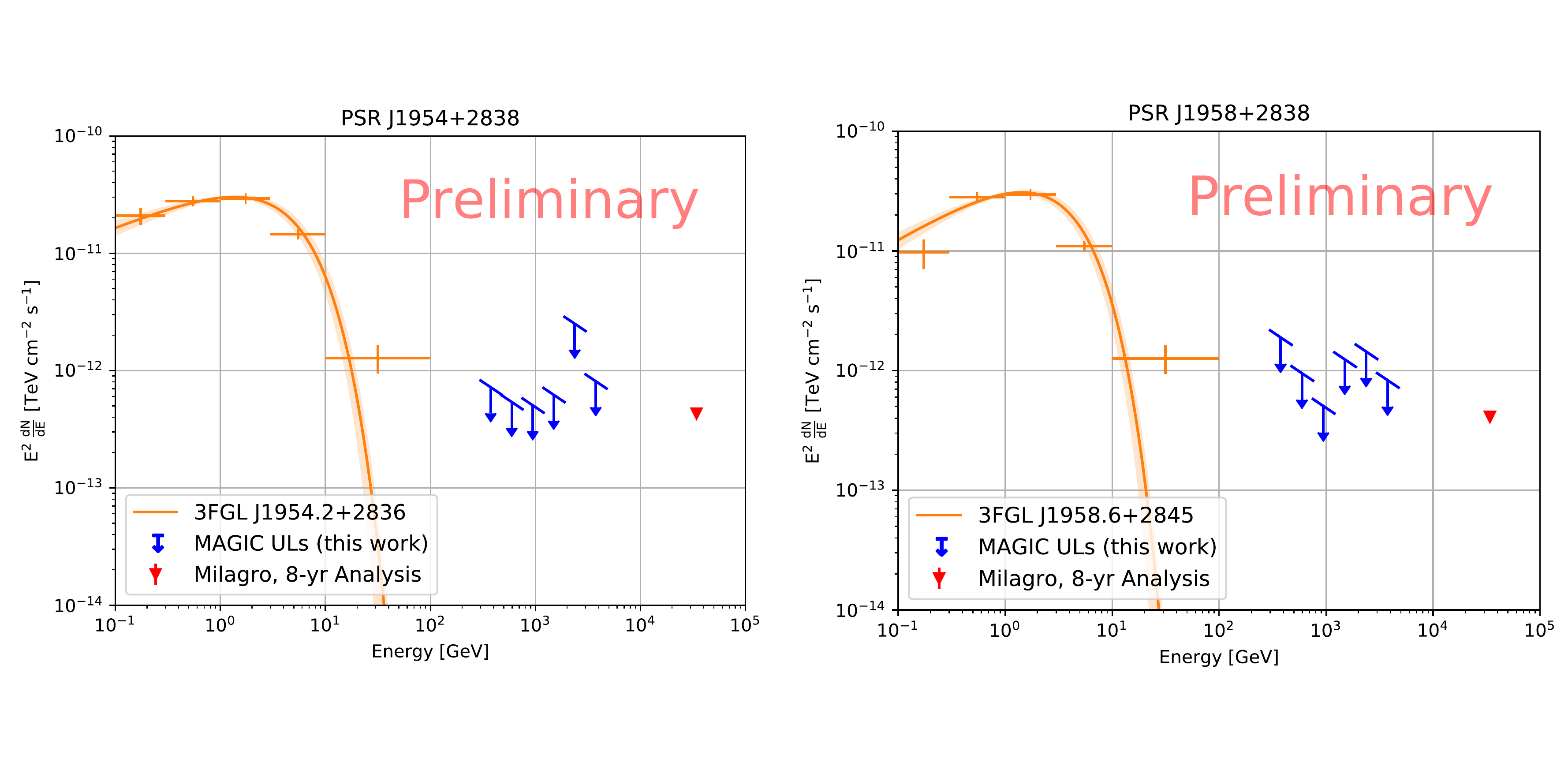}
		\caption{SED for the observation on PSR J1954+2838 and PSR J1958+2845. Results for the \textit{Fermi}-LAT pulsar as well as the Milagro re-analysis are shown, taken from \cite{1FGL} and \cite{MilagroonBrigthFermisources}, respectively.}
		\label{fig:SED_J1954_J1958}
\end{figure*}
\end{center}

The corresponding integral ULs for energies above 300 GeV assuming a power-law distribution with photon index $\Gamma=2.6$ are $1.1 \times 10^{-12}$ ph cm$^{-2}$s$^{-1}$  and $2.5 \times 10^{-12}$ ph cm$^{-2}$s$^{-1}$  ($\sim 2.6$\% CU) for PSR J1954+2838 and PSR J1958+2845, respectively. Differential ULs are listed in Table \ref{tab:difULs_J1954J1958} and depicted in the spectral energy distribution (SED) shown in Figure \ref{fig:SED_J1954_J1958}.

\begin{center}
\begin{table*}[h]
\caption{MAGIC 95\% CL differential flux ULs for PSR J1954+2838 and PSR J1958+2845 assuming a power-law spectrum with spectral index of~$\Gamma=2.6$.} 
\hfill{}
\label{tab:difULs_J1954J1958}      
	\begin{tabular}{ l  c  c}
		\hline
		\hline
		Energy range & Differential ULs & Differential ULs \\
		&  for PSR J1954+2838 & for PSR J1958+2845\\		
		\multicolumn{1}{c }{[GeV]} &[$\times10^{-13}$ TeV$^{-1}$cm$^{-2}$s$^{-1}$] & [$\times10^{-13}$ TeV$^{-1}$cm$^{-2}$s$^{-1}$] \\
		\hline
		300 -- 475.5 & 51.2 &  134.7\\
		475.5 -- 753.6 & 15.3 & 26.9\\
		753.6 -- 1194.3 & 5.7 & 5.7\\
		1194.3 -- 1892.9 &  2.8 & 5.6 \\
		1892.9 -- 3000 & 4.5 & 2.6\\
 	   	3000 -- 4754.7 & 0.6 & 0.6\\
		\hline
	\end{tabular}
	\hfill{}
\end{table*}
\end{center}

\section{Discussion and conclusions}
\begin{center}
\begin{figure*}
\centering
		\includegraphics[width=.6\linewidth]{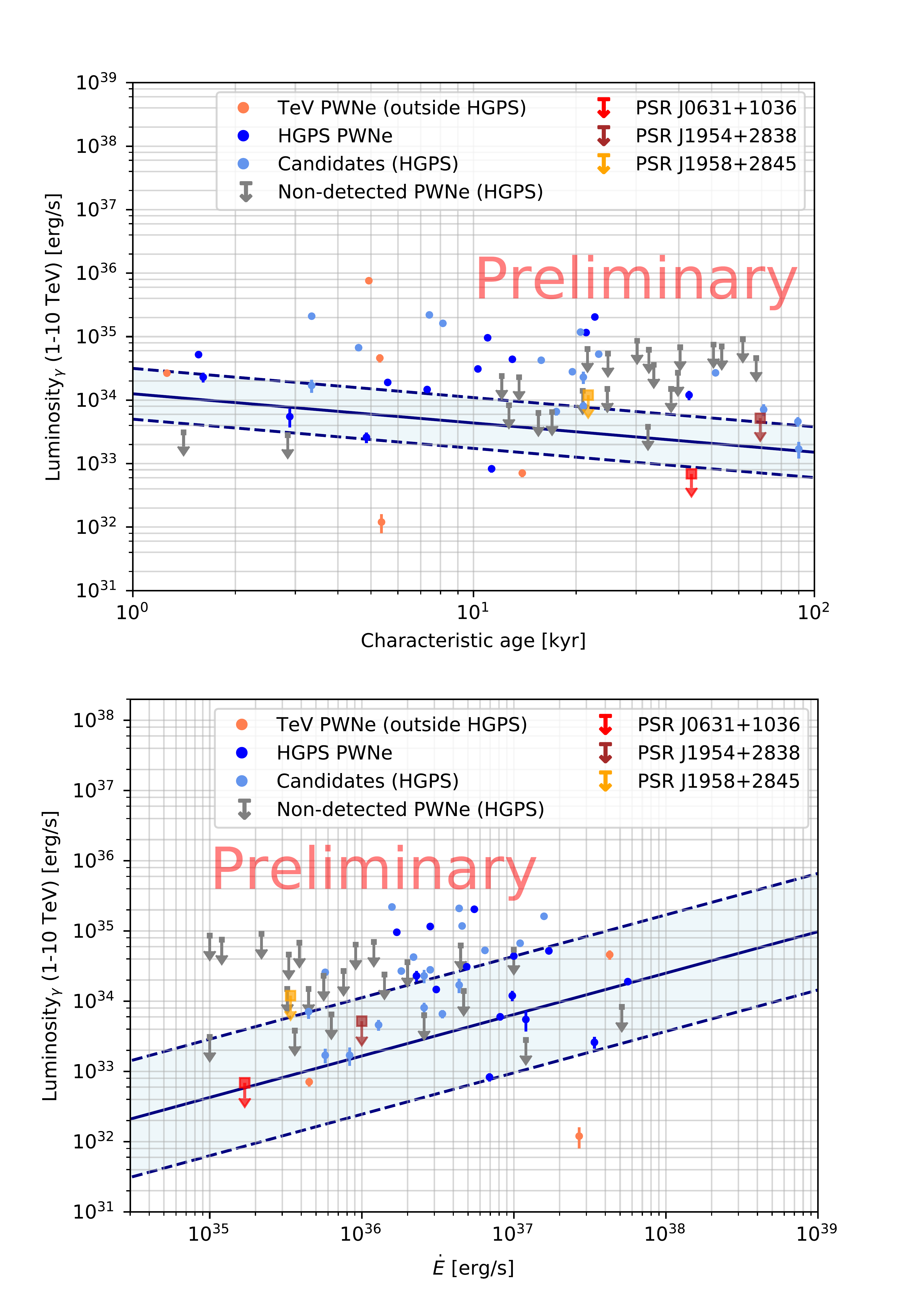}
		\caption{TeV luminosity (1--10 TeV) with respect to the characteristic age (\textit{top}) and the spin-down power of the pulsar (\textit{bottom}). The candidates included in this project are marked with squares, while external PWNe are shown with circles. In the latter, detected PWN from inside and outside of the HGPS, candidates and non-detected nebulae from it are included. The fit obtained in the study of PWN by \cite{StefanPWNstudy} is depicted as a blue band.}
		\label{fig:lum_relations}
\end{figure*}
\end{center}

Despite observing promising candidates for emitting VHE gamma rays based on observational criterion, no detection was achieved for any PWN, even though all of them, except for PSR J1958+2845, were observed for a large amount of hours. Given the location of these PWNe, at the outer parts of the Galaxy, a non-detection could be explained if different behaviors are found in the MAGIC candidates with respect to those shown by detected PWNe located mostly in the inner regions. Our results, along with all detected PWNe (inside and outside of the HGPS), HGPS candidates and the ULs obtained for the undetected HGPS PWNe (see \cite{StefanPWNstudy}) are shown in Figure \ref{fig:lum_relations}, where the PWN luminosity between 1--10 TeV is plotted versus the characteristic age and spin-down power of the hosted pulsars. The luminosity of the MAGIC PWNe are also quoted in Table \ref{tab:efficiency}. MAGIC results are in agreement with the fit obtained by \cite{StefanPWNstudy} using detected TeV PWNe and ULs. Therefore, it confirms the initial assumption that gamma-ray emission is expected from these candidates given their features.

\begin{table*}[ht!]
\begin{center}
\caption{\textit{From left to right:} Integral UL above 300 GeV, UL on the TeV luminosity (1--10 TeV), efficiency converting rotational energy into TeV gamma rays ($L_{\gamma,1-10 TeV}/\dot{E}$). To compute the luminosity, the values from the Distance column in Table \protect\ref{tab:PSR_features} were used. } 
\label{tab:efficiency}      
	\begin{tabular}{ c  c  c c}
		\hline
		\hline
		Name & Integral UL & $L_{\gamma,1-10 TeV}$ & $\xi$ \T \\
		& [$\times 10^{-12}$ ph cm$^{-2}$ s$^{-1}$]  & [$\times 10^{33} $erg s$^{-1}$]  &\\
		\hline
	     \multicolumn{1}{c}{PSR J0631+1036} & 0.6 & 0.69 & $4.1 \times 10^{-3}$ \T \\
	     \multicolumn{1}{c}{PSR J1954+2838} & 1.1 & 5.2 & $5.2 \times 10^{-3}$ \T \\
	     \multicolumn{1}{c}{PSR J1958+2845} & 2.5 & 12. & $3.5 \times 10^{-2}$ \T \\
		\hline	
	\end{tabular}
	\end{center}
\end{table*}

Possible reasons for the non-detection can be either a larger extension than expected from these sources or lower density photon fields. Both scenarios are currently being investigated and their results will be set in context with other PWNe studied by MAGIC.

\section{Acknowledgement}
We would like to thank the IAC for the excellent working conditions at the ORM in La Palma. We acknowledge the financial support of the German BMBF, DFG and MPG, the Italian INFN and INAF, the Swiss National Fund SNF, the European ERDF, the Spanish MINECO, the Japanese JSPS and MEXT, the Croatian CSF, and the Polish MNiSzW.


\begin{thebibliography}{99}
\bibitem{GaenslerSlane2006} Gaensler,~B.~M. et al., 2006, ARA\&A, 44, 17
\bibitem{StefanPWNstudy} H. E. S. S. Collaboration, Abdalla,~H., et al. 2017, ArXiv e-prints [1702.08280]
\bibitem{MAGIC_Crab} Aleksic,~J. et al., 2015, Journal of High Energy Astrophysics, 5, 30
\bibitem{Ruben_3C58} Aleksic, J.  et al. 2014, A\&A, 567, L8
\bibitem{ATNF_catalog} Manchester, R. N., Hobbs, G. B., Teoh, A., \& Hobbs, M. 2005, AJ, 129, 1993
\bibitem{PseudoDistance} Saz Parkinson, P. M.  et al. 2010, ApJ, 725, 571
\bibitem{TianLeahy2006} Tian, W. W. \& Leahy, D. A. 2006, A\&A, 455, 1053
\bibitem{Performancepaper} Aleksic, J. et al. 2016, Astroparticle Physics, 72, 76
\bibitem{Zanin2013} Zanin, R., Carmona, E., Sitarek, J., Colin, P., \& Frantzen, K. 2013, in Proc. of the 33st International Cosmic Ray Conference, Rio de Janeiro, Brasil
\bibitem{LiMa1983} Li, T.-P. \& Ma, Y.-Q. 1983, ApJ, 272, 317
\bibitem{Rolke2005} Rolke, W. A., L\'opez, A. M., \& Conrad, J. 2005, Nuclear Instruments and Methods in Physics Research A, 551, 493
\bibitem{MoonPerformance} MAGIC Collaboration et al., 2017, preprint, (arXiv:1704.00906)
\bibitem{MilagroonBrigthFermisources} Abdo, A. A.  et al. 2009b, ApJ, 700, L127
\bibitem{2HAWCcatalog} Abeysekara A. U., et al., 2017, preprint, (arXiv:1702.02992)
\bibitem{1FGL} Abdo, A. A. et al. 2010, ApJS, 188, 405
\bibitem{MAGICJ1954J1958_2010} Aleksic, J. et al. 2010, ApJ, 725, 1629
\end{thebibliography}
\end{document}